\documentclass{emulateapj}

\usepackage{color}
\newcommand{\arcsecs}{\hbox{$^{\prime\prime}$}}
\def\etal{{\it et al.}}

\begin{document}

\title{Small-Scale Structuring Of Ellerman Bombs at Solar Limb}
\author{C. J. {Nelson}\altaffilmark{1,2}, E. M. Scullion\altaffilmark{3,4}, J. G. Doyle\altaffilmark{1}, N. Freij\altaffilmark{2}, R. Erd\'elyi\altaffilmark{2}.}

\shortauthors{Nelson \it{et al.}} 

\altaffiltext{1} {Armagh Observatory, College Hill, Armagh, N. Ireland, UK, BT61 9DG.}
\altaffiltext{2} {Solar Physics and Space Plasma Research Centre, University of Sheffield, Hicks Building, Hounsfield Road, Sheffield, UK, S3 7RH.}
\altaffiltext{3} {Institute of Theoretical Astrophysics, University of Oslo, 0371 Oslo, Norway}
\altaffiltext{4} {Astrophysics Research Group, School of Physics, SNIAM, Trinity College Dublin, Dublin 2, Ireland}

\begin{abstract}
Ellerman bombs (EBs) have been widely studied in recent years due to their dynamic, explosive nature and apparent links to the underlying photospheric magnetic field implying that they may be formed by magnetic reconnection in the photosphere. Despite a plethora of researches discussing the morphologies of EBs, there has been a limited investigation of how these events appear at the limb, specifically, whether they manifest as vertical extensions away from the disc. In this article, we make use of high-resolution, high-cadence observations of an Active Region (AR) at the solar limb, collected by the {\it CRisp Imaging SpectroPolarimeter} (CRISP) instrument, to identify EBs and infer their physical properties. The upper atmosphere is also probed using the Solar Dynamic Observatory's {\it{Atmospheric Imaging Assembly}} (SDO/AIA). We analyse $22$ EB events evident within these data, finding that $20$ appear to follow a parabolic path away from the solar surface at an average speed of $9$ km s$^{-1}$, extending away from their source by $580$ km, before retreating back at a similar speed. These results show strong evidence of vertical motions associated with EBs, possibly explaining the dynamical `flaring' (changing in area and intensity) observed in on-disc events. Two in-depth case studies are also presented which highlight the unique dynamical nature of EBs within the lower solar atmosphere. The viewing angle of these observations allows for a direct linkage between these EBs and other small-scale events in the H$\alpha$ line wings, including a potential flux emergence scenario. The findings presented here suggest that EBs could have a wider-reaching influence on the solar atmosphere than previously thought, as we reveal a direct linkage between EBs and an emerging small-scale loop, and other near-by small-scale explosive events. However, as previous research found, these extensions do not appear to impact upon the H$\alpha$ line core, and are not observed by the SDO/AIA EUV filters.
\end{abstract}


\section{Introduction}
The solar atmosphere is a complex and dynamic environment, filled with a myriad of structures, ranging from large-scale coronal loops and prominences to small-scale granules and photospheric magnetic bright points (MBPs). With the increased resolution and coverage of both ground-based and space-borne instrumentation in recent years, it has become possible to observe and analyse a wider range of solar phenomena in greater detail. As certain ground-based instruments, such as the {\it CRisp Imaging SpectroPolarimeter} (CRISP; see \citealt{Scharmer06}, \citealt{Scharmer08}), are capable of resolving the lower solar atmosphere on spatial scales close to $90$ km, a wide variety of small-scale events have been discussed, specifically in terms of how they interact with the wider environment.

\begin{figure*}
\center
\includegraphics[scale=0.7]{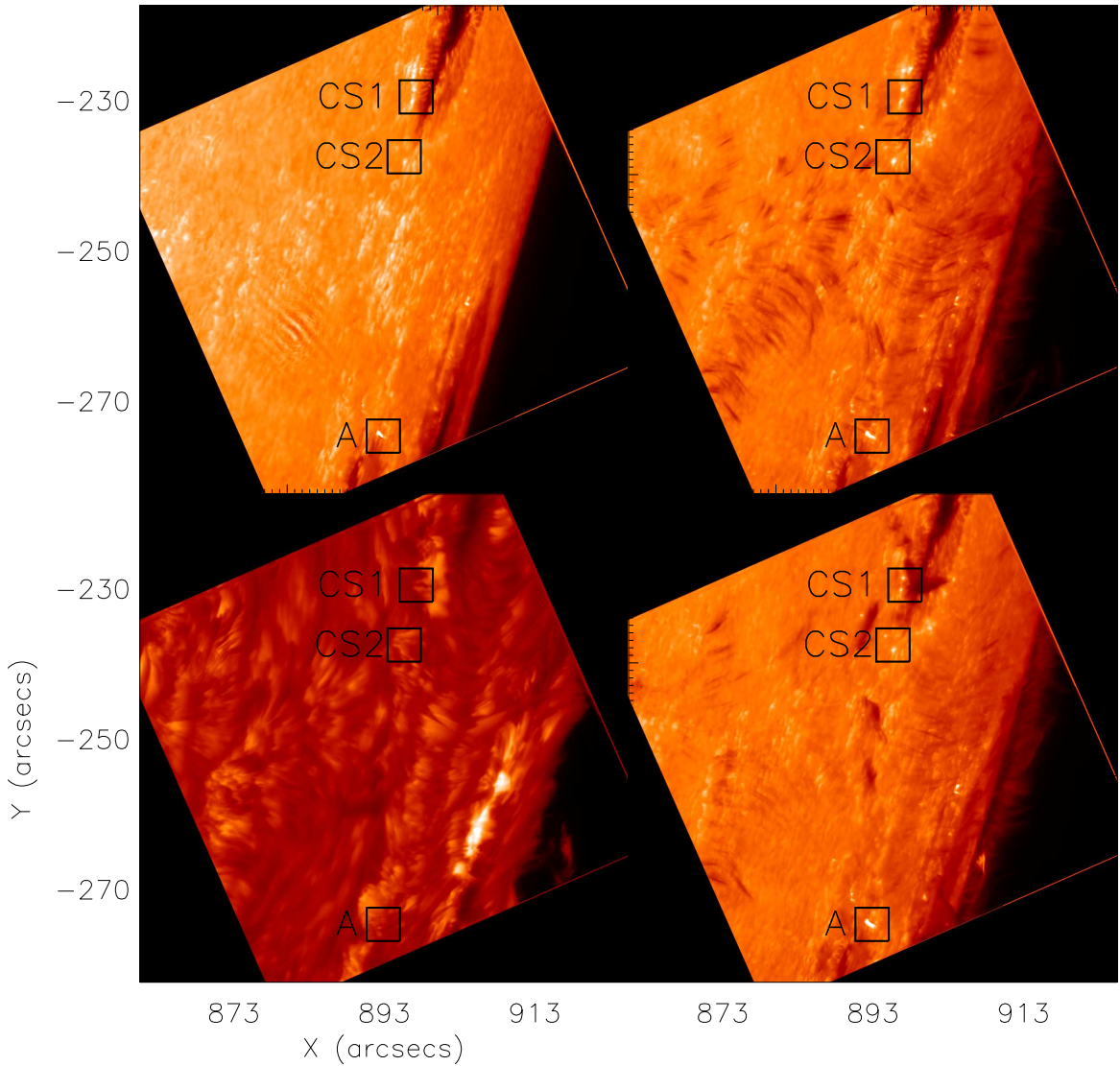}
\caption{The FOV of AR 11506 analysed within this article (corrected for local instrumental wavelength shifts), sampled at four positions within the H$\alpha$ line scan: The far blue wing (approximately $-1.8$ \AA; top left), the near blue wing ($-0.95$ \AA; top right), the H$\alpha$ line core ($0$ \AA; bottom left), and the red wing ($+0.95$ \AA; bottom right). The black boxes in each image indicate the regions of interest analysed in the following Sections. A known artifact of the image reconstruction process is visible in the far blue wing at ($883$, $-260$).}
\label{fig1}
\end{figure*}

Ellerman bombs (often shortened to EBs within the literature) are one example of small-scale events observed in the lower solar atmosphere. Widely identified as brightenings inferred from $0.5$ - $5$ \AA\ into the wings of the H$\alpha$ line profile, EBs often form co-spatially with regions of strong magnetic field, specifically in mixed polarity regions (see, for example, \citealt{Pariat04}, \citealt{Vissers13}). First observed by \citet{Ellerman17}, these small-scale events are reported to have diameters ranging from around $300$ km to $750$ km and lifetimes often less than $20$ minutes (as estimated by, {\it e.g.}, \citealt{Zachariadis87}, \citealt{Watanabe11}, \citealt{Nelson13}), meaning they are observed at the lower limits of current instrumentational capabilities. 

Recently, \citet{Nelson13} presented an analysis, using a thresholding technique, of small-scale regions of intense brightening in the H$\alpha$ line wings, inferred using the {\it Interferometric BIdimensional Spectrometer} (IBIS; \citealt{Cavallini06}) situated at the Dunn Solar Telescope (DST), and found a dynamic behaviour within many events. It was concluded that many of these small events may be EBs and that higher-resolution data should decrease the average observed size of EBs. However, as was discussed by \citet{Rutten13}, the strong network observable within the H$\alpha$ wings may also influence the thresholding technique meaning that a proportion of the less dynamic small events could be purely network brightenings and more likely associated with magnetic bright points (MBPs).

Due to both the dynamic nature of EBs and their co-spatial formation with strong magnetic fields, it is widely hypothesised that these events are observational evidence of magnetic reconnection in the upper photosphere. \citet{Georgoulis02} presented an analysis of data collected during the Flare Genesis Experiment (see \citealt{Rust00}), inferring three cartoon topologies which could lead to magnetic reconnection within the lower solar atmosphere. Two of these correspond to the formation of $\Omega$ and $\cup$ shaped topologies due to flows within the lower atmosphere, thought to be consistent with small bi-polar regions observed in magnetogram data. By applying a linear force-free extrapolation pioneered by \citet{Demoulin97}, \citet{Pariat04, Pariat07} discussed the nature of the photospheric magnetic fields close to EB events. It was found that approximately $87$\% of EBs formed co-spatially with $\cup$-shaped magnetic topologies and, hence, concluded that a serpentine flux emergence model could lead to EB formation. Further to this, \citet{Matsumoto08} presented observations of flows co-spatial to EBs potentially supporting the magnetic reconnection model. It was found that down- and up-flows in the photosphere and chromosphere existed, respectively, suggesting bi-dimensional plasma ejection by an explosive event, possibly sourced in the upper photosphere. Interestingly, during a detailed analysis of three suspected EBs by \citet{Bello13}, one event was observed to penetrate through the chromospheric canopy into the H$\alpha$ line core, indicating that large vertical flows within these events can sometimes occur.

\begin{figure*}
\center
\includegraphics[scale=0.7]{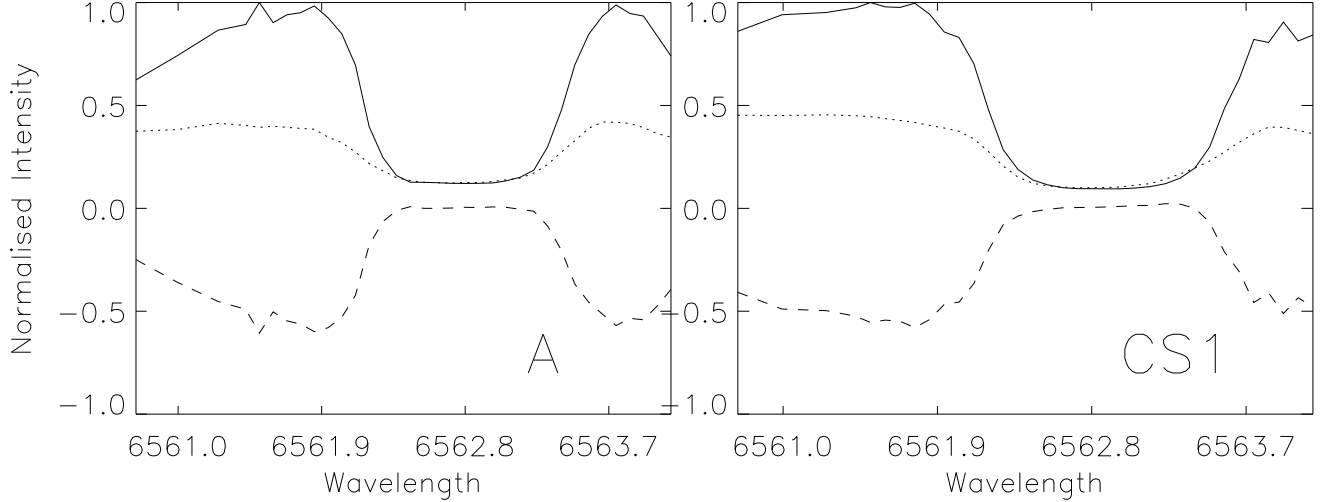}
\caption{Normalised line profiles of two representative EBs compared to the background intensity of the nearby quiet Sun. The EB line profiles (solid lines) for (left) box `A' and (right) box `CS1' in Fig.~\ref{fig1} compared to the quiet Sun (dotted line). The dashed line shows the inverted (for clarity) difference in normalised intensity between the quiet Sun and EB profiles. The heightened line wings of the EB line profile are representative of other events within these data.}
\label{fig2}
\end{figure*}

A small number of studies exist which discuss observations of EBs located close to the limb of the Sun. \citet{Roy73} presented observations of EBs for two distinct Active Regions (ARs), one close to the disc centre and one towards the limb. It was found that the apparent lifetime of EBs was shorter when observed at the limb and, importantly, that vertical extensions were evident. This work was expanded upon by \citet{Kurokawa82}, who analysed a large group of sunspots at the limb. These observations confirmed vertical extensions of EBs and provided the first quantification of lengths, at approximately $800$ km, and widths, below $450$ km for $80$\% of EBs. It is interesting to note, that this estimate of width agrees well with the measured diameters of EBs on the disc by, for example, \citet{Georgoulis02} and \citet{Nelson13}. More recently, using high-resolution H$\alpha$ data, \citet{Watanabe11} discussed the small-scale dynamics of EBs at a viewing angle of $\mu=0.67$, finding evidence of rapid increases in area, intensity, and vertical extensions. These `flaring', morphological changes were identified as evidence of a high-energy driver within the lower atmosphere, specifically, magnetic reconnection. It is clear that further analysis of events observed close to the limb at high inclination angles, as presented here, could provide interesting and useful results about the physical nature of EBs.

As well as observations, numerical methods have also been exploited to analyse the physical properties of EBs. \citet{Fang06} presented a semi-empirical model, finding that increased temperature in the lower atmosphere could lead to H$\alpha$ line profiles with increased intensity in the line wings, analogous to EBs. Numerical simulations, using the Coordinate Astronomical Numerical Softwares (CANS) code, were presented in two-  and three-dimensions by \citet{Isobe07} and \citet{Archontis09}, respectively, who found that flux emergence from below the photosphere could lead to $\cup$-shaped magnetic topologies and associated brightening events. More recently, a study was conducted by \citet{Nelson13b} who found H$\alpha$ wing brightenings, analogous to EBs, co-spatial with magnetic reconnection events within a MPS/University of Chicago Radiative MHD (MURaM) simulation box. This reconnection occured in the upper photosphere and led to increased temperatures which explained the enhanced H$\alpha$ wings, as hypothesised by \citet{Fang06}. Flows were also found, analogous to those observed by \citet{Matsumoto08}, around the reconnection site. Interestingly, both the observations and the simulations showed \ion{Fe}{1} $6302.5$\AA\ line core brightenings, which have been widely associated with magnetic reconnection (see, for example, \citealt{shelyag_lines}). 

In this article, we analyse both the morphology of EBs observed at the limb and any potential relationship between these events and the surrounding plasma. We structure our work as follows: In Section 2 we discuss the data analysed in this article; Section 3 presents our results, including a statistical analysis of EBs within these data and two individual case-studies. We discuss the implications of our findings in Section 4.

\begin{figure*}
\center
\includegraphics[scale=0.18]{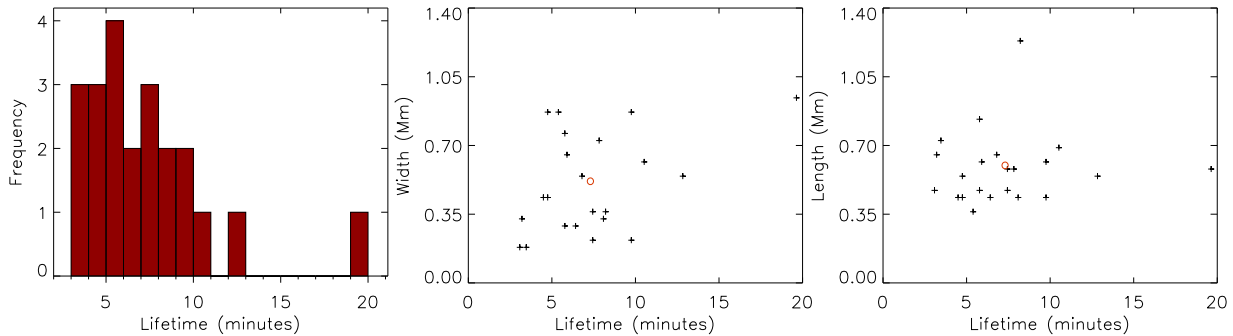}
\caption{Basic statistical properties of EBs. (Left) Lifetime of each of the $22$ identified events, with all but three events existing for less than $10$ minutes. (Centre) Distribution of peak width vs lifetime. (Right) Same as centre but for length. The red circles indicate the mean of both variables for each plot.}
\label{fig3}
\end{figure*}

\section{Observations}

The ground-based data analysed in this research were collected using the CRISP instrument attached to the Swedish $1$-m Solar Telescope (SST; \citealt{Scharmer03}) during a period of good seeing on the $21$st June 2012. A large field-of-view (FOV), situated around AR 11506 ({\it xc}=893\arcsecs, {\it yc}=-250\arcsecs\ with respect to the disc centre), containing three sunspots was selected for observations during the period from 7:15:09 UT until 7:48:25 UT. Halpha line scans, sampling 35 evenly spaced spectral positions (each with eight repetitions) between -2 \AA\ and +1.2 \AA\ from the line core (6562.8 \AA) were obtained, and further processed using the Multi-Object Multi-Frame Blind Deconvolution (MOMFBD; \citealt{vannoort05}) image restoration method. We followed the standard procedures in the reduction pipeline for CRISP data (\citealt{Rodriguez14}) which includes the post-MOMFBD correction for differential stretching suggested by \citet{Henriques12} (also see \citealt{Sekse12} for more details). Following this reduction (which included dark- and flat-fielding), the pixel size of these data was $0.059\arcsec$, which corresponds to approximately $43$ km in a transverse scale (future distance measurements within this article will discuss distances measured using this transverse scale), and the temporal cadence was approximately $7.7$ seconds, hence allowing for a detailed analysis of EB events. To conduct the analysis presented here, we exploit the excellent widget-based CRISPEX package (see \citealt{Vissers12}).

We also make use of data taken by the Solar Dynamics Observatory's {\it{Atmospheric Imaging Assembly}} (SDO/AIA; see \citealt{Lemen12}) instrument. These data image the entire solar disc within the UV spectrum using filters around $1600$ \AA\ and $1700$ \AA. These data have effective spatial and temporal resolutions of around $1000$ km and $48$ seconds, respectively. The outer layer of the solar atmosphere is also observed using a number of EUV filters; however, in the analysis presented here, we show only the $304$ \AA\ filter. Each EUV image has a spatial resolution of approximately $1000$ km and a temporal resolution of $24$ seconds. 

In Fig.~\ref{fig1}, we present an overview of the FOV analysed within this article. Clockwise from the top left image, we plot the array returned by the narrow-band CRISP filter for each of $-1.8$ \AA, $-0.95$ \AA, $+0.95$ \AA, and $0$ \AA\ at approximately 7:36:20 UT. Within this FOV, sit three sunspots (two in the southern part of the FOV and one in the northern section) as well as a large plage region. It is around these sunspots that all EBs occur that are of interest for this analysis. What is immediately apparent from these images, is that the underlying photosphere is obscured by absorption resulting from the chromospheric material in the H$\alpha$ line core in the bottom left image. The complex fibril structures observed in the line core appear to be present higher in the atmosphere, hence, potentially obscuring some of the vertical extent of the EBs in this dataset. We overlay three boxes on all images highlighting the three EBs analysed in detail.

\section{Results}

\subsection{EB statistics}

Within the H$\alpha$ line profile, EBs are easily identified as increases in intensity observed in images around $0.5$ \AA\ either side of the line core. Modern imaging-spectrometers, such as the CRISP instrument, are able to provide fast wavelength tuning between each line position included in a line scan allowing confident identification of EBs. Here, we define EBs as events which show both intensity increases which are greater than $1.5$ times the intensity of the nearby quiet Sun in the wings of the H$\alpha$ line profile and also a dynamic, explosive nature evidenced by imaging data. Fragmenting EBs observed within these data are classified as being a single event. If an event dies completely and does not recur for five frames, any new co-spatial brightening is classified as a new EB event. By employing these guidelines, we remove the influence of network bright points such as those situated at {\it xc}=893\arcsecs, {\it yc}=-250\arcsecs, which have a consistently lower line wing intensity throughout these observations, as compared with EB wing intensity excess. Overall, we confidently identify $22$ EB events within these observations. Three regions which contain EBs during these observations are highlighted in Fig.~\ref{fig1} by black boxes for further analysis. 

In Fig.~\ref{fig2}, we plot normalised line profiles for two of the representative EB events highlighted in Fig.~\ref{fig1} (solid lines; scaled to the maximum intensity of the EB profile). The significant intensity increases within these events are evident when compared to the local quiet Sun (dotted lines; also scaled to the maximum intensity of the EB profile). To highlight the percentage increase in intensity, we also plot the inverted (for visual ease) difference between the quiet Sun and the EB events (dashed lines). The difference between line-wing intensities peaks at over $-0.5$ indicating a doubling of the intensity from the quiet Sun for these EB events in their respective frames. Such gradients between EBs and the background atmosphere are not observed in every frame as the intensity of individual events appears to vary on timescales of seconds (as was discussed by \citealt{Qiu00}). We note that an acceptable thresholding value is highly dependent on a number of factors such as the instrumentation, data processing techniques, and the seeing at the time of the observations. 

After the identification of all apparent EBs in the data, each event was carefully analysed to determine its lifetime and area. As the definitions between EBs and the background are strong in these data (as is shown in Fig.~\ref{fig2}), the estimation of the lifetime was easily completed by analysing the evolution of each EB through time. The initial and final frames of each event were identified visually by studying the imaging data which show the evolution of these events clearly. Overall, the average lifetime of EBs in these data was found to be approximately $7$ minutes, comparable to previous researches by, {\it e.g.}, \citet{Roy73}, \citet{Watanabe11}, and \citet{Nelson13}. The shortest and longest-lived events, respectively, were three minute and around $20$ minutes. The distribution of lifetimes within these data is plotted in Fig.~\ref{fig3}a.

\begin{figure}
\center
\includegraphics[scale=0.37]{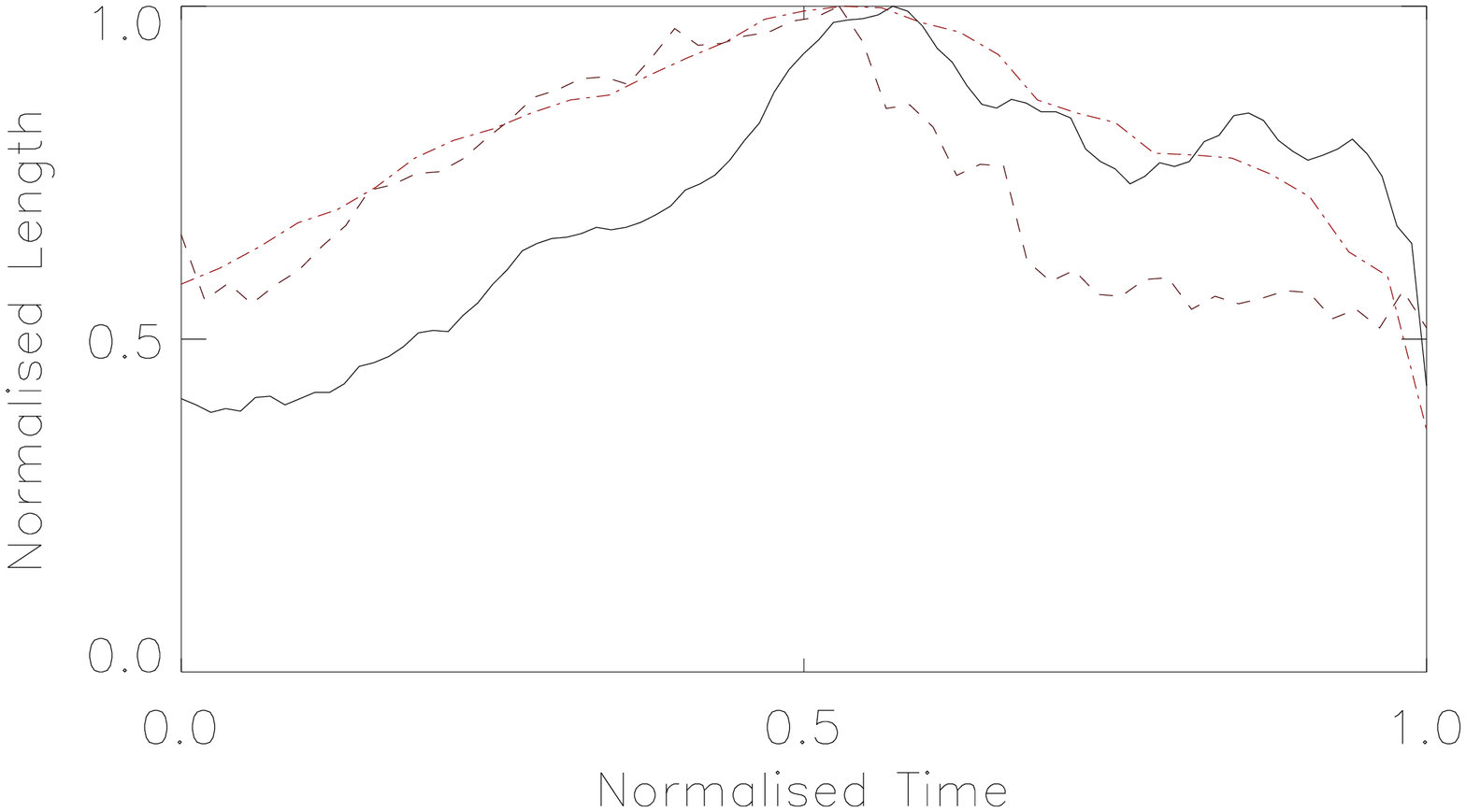}
\includegraphics[scale=0.37]{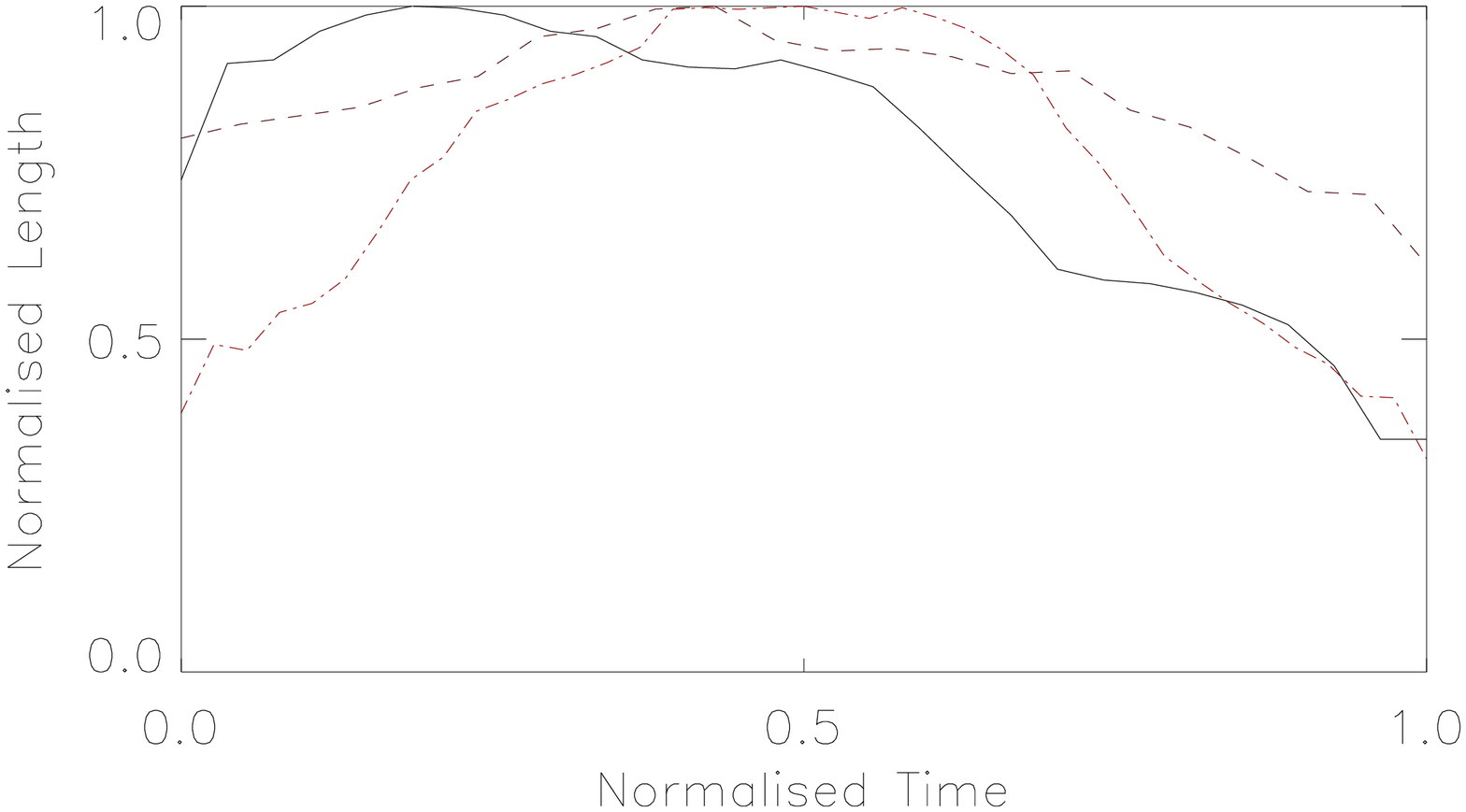}
\caption{The smoothed projection profiles for the tips (normalised against the peak length of each event) of the three EB segments analysed in Case Study: I (top) and three other independent EBs (bottom). Each event is plotted through its full lifetime. The top panel highlights the parabolic evolutions of each small segment, indicative of a repetitive driver. These profiles are similar to the majority of EB events, represented by the dashed and dot-dashed lines in the bottom panel.}
\label{fig4}
\end{figure}

\begin{figure*}
\center
\includegraphics[scale=0.55]{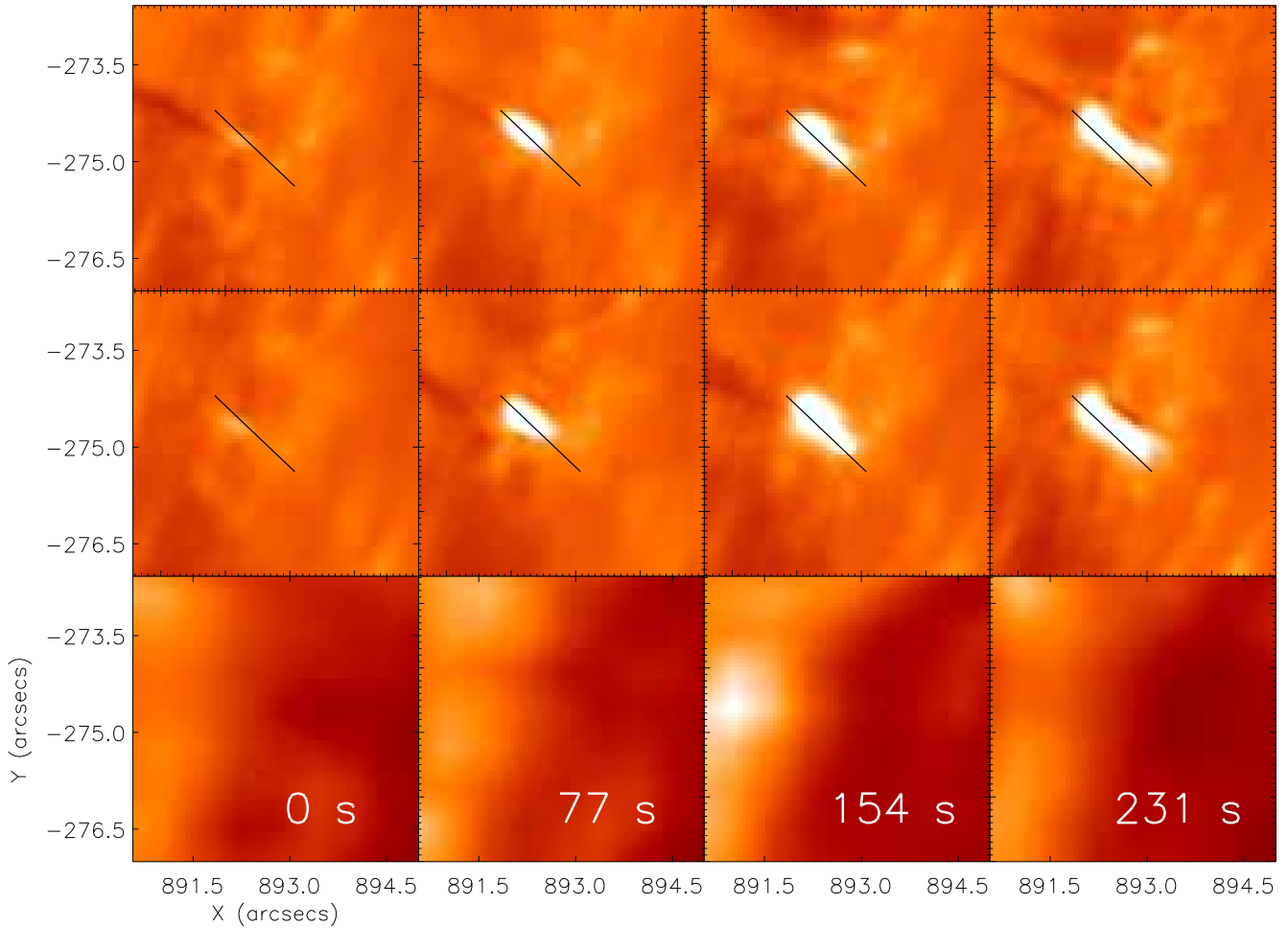}
\caption{Evolution of EB event `A' (from Fig.~\ref{fig1}) in the blue (top) and red (middle) wings of the H$\alpha$ line profile with the co-spatial SDO/AIA $304$ \AA\ filter (bottom). The EB event appears in both wings simultaneously from an apparent footpoint before extending along a constant axis (identified by the black line of length $1200$ km) to its peak length. After the fourth frame, it then fades back along the same axis. The initial images of this figure were taken at 17:32:30 UT and each subsequent image $77$ seconds later.}
\label{fig5}
\end{figure*}

We plot both the peak width and peak length of each EB event against lifetime in Fig.~\ref{fig3}b and Fig.~\ref{fig3}c. It is apparent (although not significantly correlated) that longer lived EBs appear to be larger (as previously discussed by \citealt{Roy73}). As the majority of EBs exhibit parabolic morphological evolutions through time (as evidenced in Fig~\ref{fig4}), it appears that the strength of the initial driver is a key variable in defining the statistical properties (such as lifetime and area) of each EB event. It should be noted that parabolic and ballistic profiles would not be discernible within these data due to their similarity in the photosphere and the spatial resolution. Therefore, we use `parabolic' as an umbrella term for both profile types. Basic energy estimates of EBs (see, for example, \citealt{Georgoulis02}) require both lifetime and area and, hence, it would appear that a correlation exists between lifetime and energy release. Future analysis with a larger statistical sample should further test this assertion.

In Fig.~\ref{fig5}, a representative EB event is plotted through its onset until it reaches its peak length. The EB appears simulataneously and co-spatially in both wings before extending away along a constant trajectory. The black lines in Fig.~\ref{fig5} indicate the path of the EB over time. It is interesting to note that all but two of the EBs analysed appear to have tips which extend and contract with parabolic profiles, however, horizontal motions within these events are also common. Each EB was carefully analysed for both vertical and horizontal motions and the results were recorded. For the parabolic EBs identified, an average vertical speed of around $9$ km s$^{-1}$ was observed (from onset to peak extension), with most events reaching higher velocities during their most explosive periods. It was found that $12$ of the EB events analysed here exhibited transverse motions, averaging at $1.5$ km s$^{-1}$ (similar to velocities observed by, for example, \citealt{Denker95}, \citealt{Nindos98}); however, several EBs had apparent motions over $3$ km s$^{-1}$. The average horizontal speed is slightly higher than previous estimates (by, {\it e.g}, \citealt{Georgoulis02}, \citealt{Watanabe11}, \citealt{Nelson13}), probably due to a small number of extremely dynamic events which are observed. We examine two EBs with significant horizontal speeds, and an apparent influence on the wider atmosphere, in detail in the following case studies.

\begin{figure*}
\center
\includegraphics[scale=0.55]{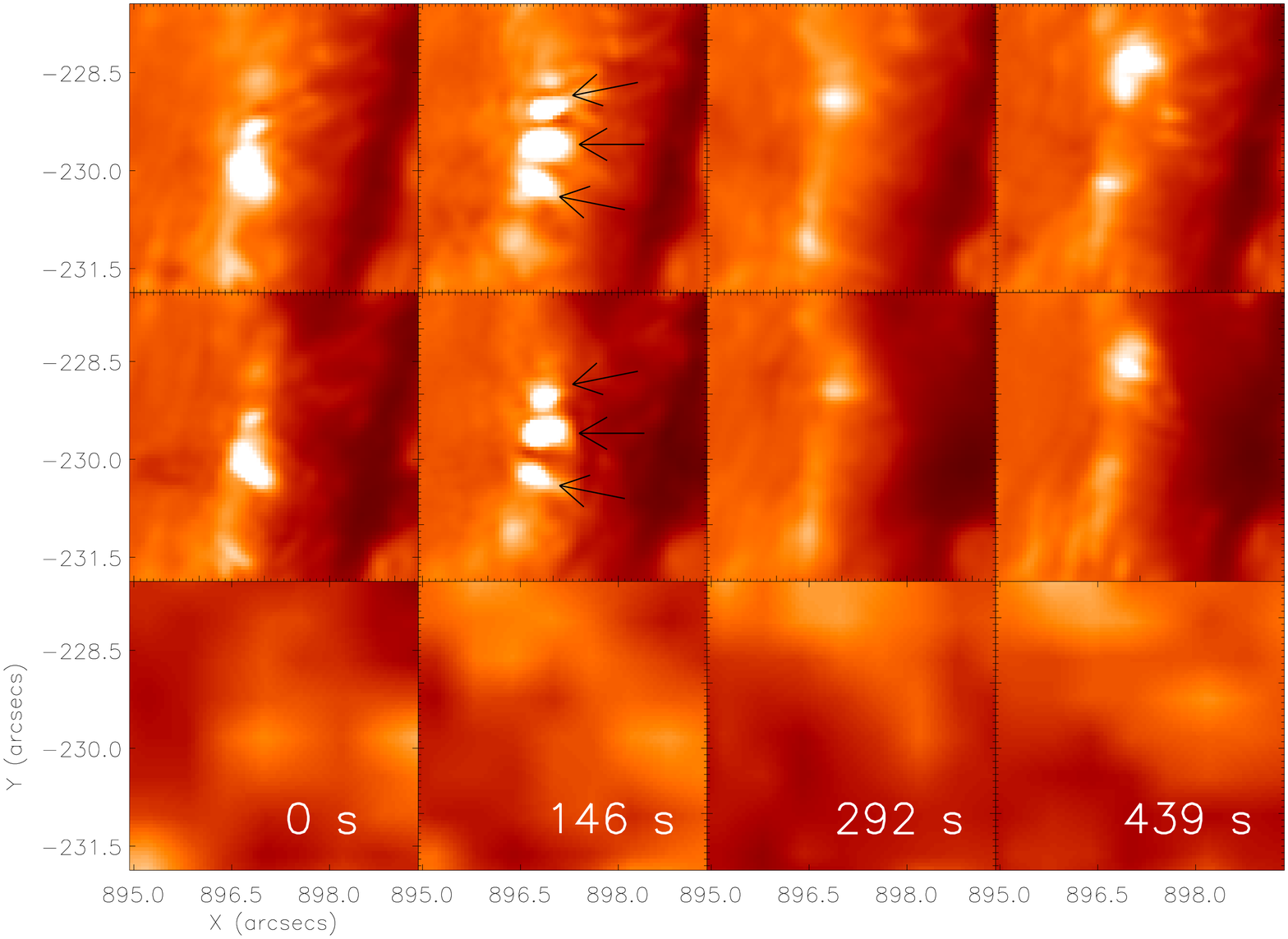}
\caption{An illustration of the propagation of the EB analysed in Case Study I for both the blue ($-1$ \AA; top) and red ($+1$ \AA; middle) wings of the H$\alpha$ line profile, as well as the SDO/AIA $304$ \AA\ filter (bottom). The almost northward propagation of this event appears to be parallel to the near-by penumbra and follows the bright track evident in the third panel. In the second panel for each wavelength, three small sub-structures are highlighted with arrows. The line wings are originally sampled at 7:29:54 UT and each subsequent image is separated by $146.3$ seconds.}
\label{fig6}
\end{figure*}

Co-spatial EUV data inferred by the SDO/AIA instrument are also analysed for each EB. As EBs consist of a vertical extension, one may expect that some signature would be observed in the upper atmosphere, however, the majority of previous studies have found no signal even within the H$\alpha$ line core (see, for example, \citealt{Zachariadis87}, \citealt{Watanabe11}, \citealt{Vissers13}). Despite recent work by \citet{Bello13} suggesting that some EBs may penetrate into the chromosphere, it is still unclear whether these events have any influence on the chromosphere and corona. The EBs analysed in this article show no influence in the upper atmosphere (although this lack of influence is to be expected within data collected at the limb), specifically within the H$\alpha$ line core, as indicated in Fig.~\ref{fig1}, or the EUV SDO/AIA filters, as plotted in Fig.~\ref{fig5} and Fig.~\ref{fig6}. Had co-spatial IRIS observations been available for this research, it would have been interesting to analyse what the overall influence of EBs is on the Transition Region plasma around 100,000 K.

As the intensity enhancement of an EB occurs in both wings of the H$\alpha$ line profile simultaneously and corresponds to apparent vertical motions, it is likely that these observables are a result of increased temperature and density within the ejected plasma compared to the surrounding atmosphere. This hypothesis agrees with simulations of EBs within the lower solar atmosphere (by, {\it e.g.}, \citealt{Fang06}; \citealt{Nelson13b}), and with observations of flows co-spatial to EB events ({\it e.g.}, \citealt{Matsumoto08}). It has been widely speculated, as previously discussed, that magnetic reconnection in the photosphere could lead to plasma ejection, hence creating density increases in the local atmosphere similar to those observed here; however, it should be noted that no magnetic field data of sufficient resolution comparable to EB cross-sections are available for comparison to the CRISP data.

\subsection{Case Study: I}

In these data, it is common that large, apparent horizontal motions are observed within EBs during their lifetimes. How these horizontal motions lead to interactions with plasma in the wider atmosphere is of specific interest and could prove key in assessing the potential influence of EBs within the solar photosphere. In previous studies, it has proved difficult to accurately link EB events with any other solar phenomena and, hence, they have been analysed as localised events. Here, we present one specific example of a region which appears to be susceptible to the formation of a number of EBs in a structured manner. EB events within this region display strong horizontal motions and appear to trigger other, similar events in different spatial locations.

In Fig.~\ref{fig6}, we plot the evolution of the northern event emphasised in Fig~\ref{fig1} with the label `CS1', with respect to time for both $6561.7$ \AA\ and $6563.83$ \AA\ ($-1.1$ \AA\ and $+1.03$ \AA\ from the line core, respectively). The first frame depicts the original EB, before spatial fracturing within this event is evident in the second frame (indicated by arrows). Each independent fracture appears to slowly propagate away from the original footpoint along the bright trail evident in the third frame. After the original EB fades for long enough such that it is deemed to have ended, a second large EB event occurs, as evidenced in the fourth frame. This rapid morphology is reminiscent of the evolution of the magnetic field simulated by, {\it e.g.}, \citet{Archontis09}, where an emerging flux rope formed in a `sea-serpent'-like manner reconnected at each individual $\cup$ to form a larger over-lying loop.

EBs have been shown to occur co-spatially with inter-granular lanes (see, for example, \citealt{Denker95}, \citealt{Nelson13}). It is possible that the bright trail which appears to guide the EBs is evidence of a localised network structure, or an inter-granular lane. On-disc observations of the H$\alpha$ line wings often include weak intensity increases, reminiscent of this trail, co-spatial to strong magnetic fields, inferred using magnetogram or $G$-band data. It is, therefore, possible that these EBs are propagating along a defined structure and, hence, that further information could be derived by analysing on-disc examples of such events. High-resolution, multi-wavelength observations close to the disc centre should be further investigated to infer whether these events are indeed guided by the magnetic field.

\begin{figure*}
\center
\includegraphics[scale=0.38]{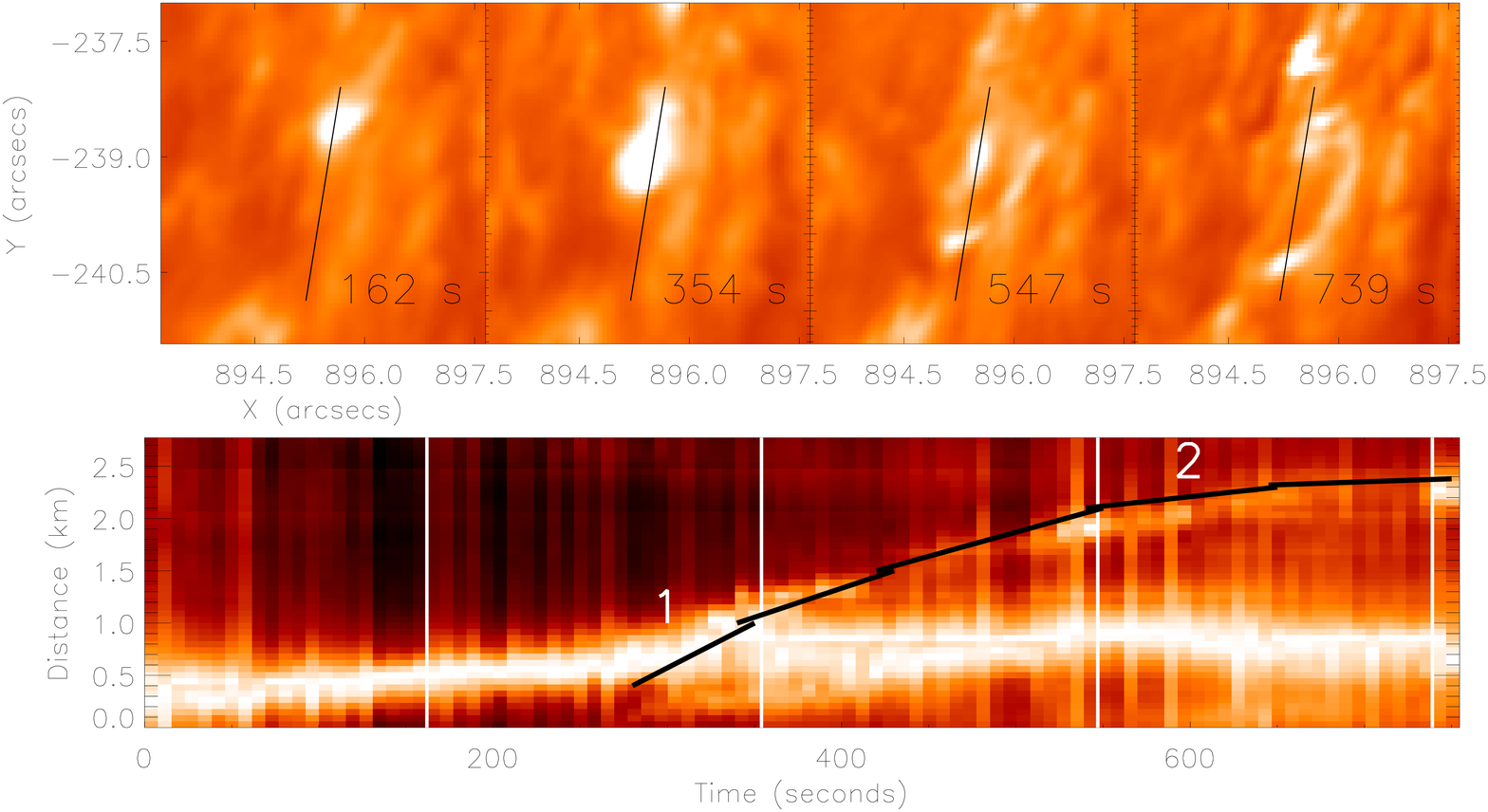}
\caption{The evolution of the EB analysed in Case Study II at $+0.8$ \AA. In the top row, four frames from this wavelength are plotted starting at 7:40:49 UT and separated by 146.3 seconds. The initial EB event is easily observed in the left-hand frame. The most dynamic horizontal motion of the event is shown in the second frame before the generation of the apparent loop is depicted in the third and fourth frames. The bottom row shows the time-distance analysis of the black slit overlaid on the top row with intensity normalised through time to remove the influence of changes in seeing. The black line indicates the speed of the event through time, highlighting the evident deceleration. The speeds of the event at the times marked by $1$ and $2$ are $6.2$ km s$^{-1}$ and $0.6$ km s$^{-1}$, respectively. White vertical lines depict the temporal position of each of the four top row plots.}
\label{fig7}
\end{figure*}

It has been widely reported that EBs both migrate (for example, \citealt{Denker95}, \citealt{Nindos98}) and appear to recur, but what has not been presented yet in such detail, is a direct link between two apparently different, and highly structured, events. This does pose several important questions which can be discussed here. For example, what process is leading to the creation of multiple vertical peaks within this individual EB? If magnetic reconnection is indeed the driver of EBs, then, are we observing a `train' of reconnection through a `sea-serpentine' morphology as simulated by \citet{Archontis09}? Thus, are sequential and apparently connected EBs a signature of a yet un-determined large-scale sub-surface process? The spatial separation between the initial and final EB event is around $2200$ km, hence, this would suggest that a single reconnection event in a unstable region could lead to a sustained energy release within the local plasma (of course, this statement also applies if another driver is the cause of these events). For a full analysis of events such as this to be completed, high-resolution magnetic field data would be required, well beyond the current capabilities of instrumentation. It is, therefore, imperative that further research be carried out using both imaging observations and state-of-the-art computational modelling.

The second important result which can be drawn from this event is that small-scale EB dynamics, as discussed by \citet{Nelson13}, are conspicuous within these data. It is inherently clear that higher-resolution data may allow further insight into the dependence, or indeed independence, of these small-scale events to close-by larger EB events. The individual parabolic profiles evident in each of the smaller-scale structures (Fig.~\ref{fig4}a) analysed in this example (analogous to the profiles observed in Fig.~\ref{fig5}) adds weight to the argument that each fragment may be formed by a separate (or a single repetitive migrating) driver. Overall, we suggest that future analysis of EBs in a wide range of datasets be conducted to assess whether a minimum EB size is determinable using modern instrumentation.

\subsection{Case Study: II}

The final case study included within this article focuses on the event highlighted in Fig.~\ref{fig1} by the box labeled `CS2'. A number of interesting morphological features are observed during the evolution of this event which further evidence the dynamical nature of EBs. This EB exhibits the most rapid apparent horizontal motions observed in this dataset (around $6$ km s$^{-1}$), which occur during an apparent splitting of the event. Such dynamics have yet to be studied in the literature and provide a potentially excellent diagnostic tool for future analysis of the driver of EBs.

The evolution of the event is depicted by the top row of Fig.~\ref{fig7} which shows information observed at $+0.8$ \AA. The original EB, initial splitting, fading, and then loop formation are visualised from left to right, respectively. The bottom row of Fig.~\ref{fig7} includes a time-distance plot for the spatial positioning indicated by the black line in the top row. The initial off-shoot appears to be as bright as the original EB event and propagates away from the formation site at a speed of $6.2$ km s$^{-1}$. This is over four times the average apparent horizontal motion speed of EBs within these data. The off-shoot continues to move away from the large EB and decelerates until it reaches a speed of around $0.6$ km s$^{-1}$. A black line is overlaid on the time-distance plot to emphasise the path of the off-shoot.

Of particular interest here is the similarity of this evolution to magnetic flux emergence events. Comparable morphological traits to these were reported by both \citet{Otsuji07} and \citet{Ortiz13}. These researches analysed events which had initial separation speeds of around $5$ km s$^{-1}$ that dropped to around $1$ km s$^{-1}$, and spatial separations of the footpoints on the order of $2200$ km - $3000$ km. Flux emergence models also commonly discuss the occurrence of bright regions at the footpoints of formed loops (by such authors as \citealt{Guglielmino08}), as observed here in the form of EBs. These brightenings have been linked to reconnection between the emerging and existing fields and could facilitate the transport of energy from the lower solar atmosphere into the corona (as found to be, for example, by \citealt{Isobe08}). Unfortunately, as this event occurs during the final frames of this dataset, we are unable to establish whether this brightening and loop structure displays the traits observed in previous studies. Interestingly, \citet{Zachariadis87} observed the occurrence of EB pairs, separated by around $3$\arcsec. It is plausible that such pairs were formed in a comparable method to that described in this subsection.

In terms of EBs, this apparent link to an observation of flux emergence could prove exciting. Magnetic flux emergence has long been discussed as a potential driver of magnetic reconnection (see, for example, \citealt{Heyvaerts77}, \citealt{Shibata92}, \citealt{Guglielmino08}) and, in particular, as a driver for EBs (suggested by, {\it e.g.}, \citealt{Georgoulis02}, \citealt{Archontis09}). We acknowledge that no co-spatial magnetic field data is available and, as such, we are unable to conclusively link this event with flux emergence, however, the similarities presented here are compelling. It should be noted by the reader that other alternatives exist to the flux emergence scenario, such as mass loading of an already existing loop. A larger-scale study of such events would be required to definitely answer this point. It is also unfortunate that these observations end co-temporally with the fourth frame of Fig.~\ref{fig7}, meaning that we are unable to analyse the full extent of this event. A variety of datasets should be analysed in the near future to further test these findings, specifically in terms of how many EBs are actually linked to examples of flux emergence.

\section{Discussion}

The results presented here support the conclusions of earlier investigations, where it has been suggested that EBs are energetic explosive events emanating from the lower solar atmosphere (see, for example, \citealt{Georgoulis02}, \citealt{Watanabe11}, \citealt{Nelson13b}). The average lifetime and spatial properties of brightenings analysed in this article ($7$ minutes and widths around $0.65$\arcsecs) are comparable to values reported by a number of authors as properties of EBs, therefore, allowing us to confidently link these near-limb events to on-disc EBs. However, we present the first limb measurements of EB lengths using state-of-the-art ground based instrumentation, finding the average height of these events to be $600$ km {(which is well below the believed height of formation for the H$\alpha$ line core of $2000$ km)}. This is slightly shorter than previous estimates by \citet{Kurokawa82}. It should be noted that a plethora of highly dynamic events within our data were observed which did not eventually reach the required intensity threshold, possibly due to a mixing of events within the line-of-sight manifesting in a single less intense line profile. Because of this, we suggest that further study of EBs at the limb with a range of datasets could provide interesting additional insights. 

We also find strong evidence of flows associated with EBs, agreeing with previous observations by, for example, \citet{Roy73} at the limb and \citet{Matsumoto08} on the disc. The tips of $20$ out of $22$ events appear to follow a parabolic path through time suggesting the occurrence of a displacement of plasma, increasing the density and temperature within a localised region, hence, leading to the enhancement of the intensity in the wings of the H$\alpha$ line profile. We suggest that this propagation of plasma is analogous to the flows observed by \citet{Nelson13b} at a simulated reconnection site where rapid cancellation of opposite polarity field occured. Unfortunately, due to the FOV of these observations being situated at the solar limb, we are unable to confidently present co-temporal, co-aligned magnetic field data to analyse with this dataset; hence, further assertions about the formation of these events elude us. We note, however, that no evidence of EBs within the H$\alpha$ line core or the SDO/AIA EUV filters is found, agreeing with previous studies which have concluded that the vertical extensions of these events may not be sufficient enough to penetrate into the chromosphere and lower corona.

The two individual case studies presented in Section $3$ highlight small-scale dynamics associated with EBs which have not previously been observed. Within the first case study, the influence of an EB on the surrounding atmosphere was analysed. A large EB event appeared (by visual inspection)  to fragment, with the small-scale pieces appearing to propagate north, away from the initial event. Each of the small-scale fragments were only around $230$ km in diameter, similar in size to the events analysed by \citet{Nelson13}. The northern-most fragment drifted to around $1500$ km from the initial position before reducing dramatically in size and fading below the threshold of $1.5$ times the background intensity. A second large EB event was, then, observed to occur at the same spatial position. This case study highlights the influence that EBs can have on the surrounding localised plasma. Other examples of the horizontal extensions of EBs are also observed within these data, however, as these events are further towards the limb, we are unable to fully resolve any potential smaller-scale structures within the larger event. 

The second case study discussed a rapid splitting of a large EB event close to a large sunspot. The main body of the ejection appeared to propagate south, away from the initial event, and continued through until the end of these observations, decelerating from around $6.2$ km s$^{-1}$ to approximately $0.6$ km s$^{-1}$. Possibly, the most interesting aspect of this example is the apparent loop formation between the two main bodies in the H$\alpha$ line wings, potentially indicating a flux emergence region (see, for example, \citealt{Otsuji07}, \citealt{Ortiz13}). Despite a significant apparent vertical extension of this loop, no evidence of any such structure within the H$\alpha$ line core was found (possibly due to the dense foreground structures in the H$\alpha$ line core obscuring any signal) suggesting that even a dynamic event, such as this example, has no initial influence on the upper chromosphere. Unfortunately, our observations end before the loop faded and we are unable to discuss the full evolution of this event. We strongly encourage that further work be carried out to fully test whether other flux emergence regions can be correlated to EBs.

Overall, we suggest that this analysis highlights both the small-scale structuring and dynamic nature of EBs. An investigation of a wide variety of these events at a range of spatial positions over the Sun would be required to fully understand how many EBs display morphologies similar to those discussed within the presented case studies. We have now addressed the importance of investigating the sub-structures of small-scale, explosive phenomena in the lower solar atmosphere which can act as important agents in triggering local instabilities in the magnetic environment of the solar surface. Such influence can be both vertically and horizontally orientated and require extensive future study.

\acknowledgements
Research at the Armagh Observatory is grant-aided by the N. Ireland Dept. of Culture, Arts and Leisure. We thank the UK Science and Technology Facilities Council for CJN's and NF's studentships, PATT T\&S support, plus support from grant ST/J001082/1. The Swedish 1-m Solar Telescope is operated on the island of La Palma by the Institute for Solar Physics of Stockholm University in the Spanish Observatorio del Roque de los Muchachos of the Instituto de Astrofísica de Canarias. We thank L. Rouppe van der Voort (Institute of Theoretical Astrophysics, University of Oslo) for advice on data reductions with MOMFBD for SST/CRISP. RE is thankful to the NSF, Hungary (OTKA, Ref. No. K83133) and acknowledges M. K\'eray for patient encouragement.

\end{document}